\documentclass[10pt,twocolumn,letterpaper]{article}

\usepackage{wacv}              

\usepackage{graphicx}
\usepackage{amsmath}
\usepackage{amssymb}
\usepackage{booktabs}
\usepackage{float}
\usepackage{xcolor}

\usepackage[pagebackref,breaklinks,colorlinks]{hyperref}

\usepackage[capitalize]{cleveref}
\crefname{section}{Sec.}{Secs.}
\Crefname{section}{Section}{Sections}
\Crefname{table}{Table}{Tables}
\crefname{table}{Tab.}{Tabs.}

\def\wacvPaperID{661} 
\def\confName{WACV}
\def\confYear{2024}
\begin{document}

\title{SCUNet++: Swin-UNet and CNN Bottleneck Hybrid Architecture with Multi-Fusion Dense Skip Connection for Pulmonary Embolism CT Image Segmentation
\thanks{This work was supported by Natural Science Foundation of Zhejiang Province (No. LY21F020015), National Natural Science Foundation of China (No. 61972121), the Open Project Program of the State Key Laboratory of CAD\&CG (No. A2304), Zhejiang University, GuangDong Basic and Applied Basic Research Foundation (No. 2022A1515110570), Innovation Teams of Youth Innovation in Science, Technology of High Education Institutions of Shandong Province (No. 2021KJ088), Education and Teaching Reform Research Project of Hangzhou Dianzi University (No. SGJYB202206) and National Undergraduate Training Program for Innovation and Entrepreneurship.

$\star$ Corresponding \hspace{0.1cm} Author
}
}

\author{Yifei Chen$^1$, Binfeng Zou$^1$, Zhaoxin Guo$^1$, Yiyu Huang$^1$, Yifan Huang$^1$, Feiwei Qin$^{\star1}$, \\
 Qinhai Li$^3$, Changmiao Wang$^{\star2}$ \\
$^1$Hangzhou Dianzi University, $^2$Shenzhen Research Institute of Big Data,\\ $^3$Second Affiliated Hospital of the Chinese University of Hong Kong \\
\tt$\{$chenyifei, qinfeiwei$\}$@hdu.edu.cn,\\
\tt$cmwangalbert$@gmail.com.\\}

\maketitle

\begin{abstract}
Pulmonary embolism (PE) is a prevalent lung disease that can lead to right ventricular hypertrophy and failure in severe cases, ranking second in severity only to myocardial infarction and sudden death. Pulmonary artery CT angiography (CTPA) is a widely used diagnostic method for PE. However, PE detection presents challenges in clinical practice due to limitations in imaging technology. CTPA can produce noises similar to PE, making confirmation of its presence time-consuming and prone to overdiagnosis. Nevertheless, the traditional segmentation method of PE can not fully consider the hierarchical structure of features, local and global spatial features of PE CT images. In this paper, we propose an automatic PE segmentation method called SCUNet++ (Swin Conv UNet++). This method incorporates multiple fusion dense skip connections between the encoder and decoder, utilizing the Swin Transformer as the encoder. And fuses features of different scales in the decoder subnetwork to compensate for spatial information loss caused by the inevitable downsampling in Swin-UNet or other state-of-the-art methods, effectively solving the above problem. We provide a theoretical analysis of this method in detail and validate it on publicly available PE CT image datasets FUMPE and CAD-PE. The experimental results indicate that our proposed method achieved a Dice similarity coefficient (DSC) of 83.47\% and a Hausdorff distance 95th percentile (HD95) of 3.83 on the FUMPE dataset, as well as a DSC of 83.42\% and an HD95 of 5.10 on the CAD-PE dataset. These findings demonstrate that our method exhibits strong performance in PE segmentation tasks, potentially enhancing the accuracy of automatic segmentation of PE and providing a powerful diagnostic tool for clinical physicians. Our source code and new FUMPE dataset are available at \href{https://github.com/JustlfC03/SCUNet-plusplus}{https://github.com/JustlfC03/SCUNet-plusplus}.
\end{abstract}

\section{Introduction}
PE is a severe and life-threatening condition caused by the dislodgment of an embolus within a blood vessel, which obstructs a pulmonary artery or its branches. The embolus size determines its classification as a central, lobar, segmental, or subsegmental pulmonary artery. Small emboli may present with minimal symptoms or mild chest tightness, while larger emboli can lead to fainting or sudden death. This is due to the thrombus obstructing the pulmonary artery, causing narrowing or blockage of the patient's blood vessel, resulting in increased pulmonary vascular resistance and elevated pulmonary artery pressure, ranking second in severity only to myocardial infarction and sudden death. The incidence of PE rises with age, and without active treatment, its mortality rate will approach 30\%. However, with timely and appropriate intervention, this rate can be reduced to 2\% to 11\% \cite{blackmon2011computer}. A significant clinical challenge is the prompt and accurate diagnosis of PE while minimizing the risks associated with false-positive diagnoses.

CTPA is the primary diagnostic method for PE. In CTPA, a contrast agent dissolves in the blood, causing the blood vessels to appear bright, while emboli do not absorb the contrast agent, resulting in dark areas that represent PE in CT images. According to radiologists' experience, the voxel values of PE typically range from -50 HU to 100 HU which represents the image brightness level \cite{buhmann2007clinical}. Radiologists use this feature to screen for PE by meticulously tracing each artery branch. However, PE diagnosis is a highly complex task for several reasons. Firstly, a patient may have hundreds or thousands of CTPA images, with each image representing a slice of the lungs. Repetitive image readings can consume significant time and effort for physicians, leading to a higher risk of misdiagnosis. Secondly, CTPA images can be affected by imaging techniques and factors such as respiratory motion artifacts, lymph nodes, and vascular branching, introducing noise interference. Finally, CTPA requires a high level of expertise, and insufficient understanding on the part of physicians can lead to delayed diagnosis and missed cases. To address these challenges, computer-aided detection (CAD) serves as an important tool for radiologists. CAD enables accurate detection and diagnosis of PE while reducing the CTPA reading time, thereby enhancing overall diagnostic efficiency.

In this paper, from a clinical perspective by learning how doctors locate PE in CT images based on its features, we propose a PE segmentation method, called SCUNet++. The method focuses on CTPA images and combines the advantages of UNet++, multiple fusion dense skip connections, Swin-Transformer attention mechanism, and Swin-UNet method, and we made some improvements to enhance the segmentation accuracy and stability of the network. The primary contributions of this paper are as follows:

\begin{enumerate}
  \item We have curated a CT image dataset annotated for PE. In collaboration with medical professionals, our team meticulously analyzed the publicly accessible FUMPE and CAD-PE datasets. These examinations revealed substantial inaccuracies and inconsistencies within the original dataset annotations. As a corrective measure, we undertook the task of re-annotating these datasets to guarantee their precision. Furthermore, to facilitate accessibility and utility for other users, we have provided a google drive downloadable link for the amended FUMPE dataset on our GitHub repository.
  
  \item We investigate the amalgamation of the UNet++ convolutional neural network featuring skip connections, the Transformer attention mechanism, and the Swin-UNet pure Transformer U-shaped neural network model. This integrated approach results in an elevation of segmentation accuracy.
  
  \item A bottleneck module that utilizes CNN blocks is introduced to address the Swin-Transformer blocks' inadequacy in extracting local spatial features from images. This strategic addition subsequently leads to an enhancement in the overall performance of the model.
\end{enumerate}

\section{Related Work}
Traditional medical image segmentation methods primarily rely on computer graphics and machine learning techniques for segmentation. More mature approaches include threshold-based image segmentation \cite{yi2012research}, regional image segmentation utilizing similarity of region features \cite{mesanovic2011automatic}, boundary segmentation employing edge detection operators \cite{anandh2016method, wei2022matr}, and contour segmentation based on curve evolution. However, traditional medical image segmentation methods predominantly rely on low-level visual features of image pixels. Owing to the substantial variation in the shape and contour of organs, the segmentation results may fluctuate or even decline as the complexity of the segmented object increases during practical applications, making it challenging to ensure adequate accuracy. Consequently, traditional medical image segmentation methods exhibit limited effectiveness in enhancing segmentation accuracy.

With the widespread application of machine learning, the application of convolutional neural networks for medical image processing has also expanded, primarily falling into two categories: classification and segmentation. 

For classification tasks, Yang \textit{et al.} \cite{yang2019two} designed a two-stage model for detecting PE using 3D CTPA images. Initially, candidate regions are proposed, followed by the extraction of vessel-aligned 3D candidate regions. False positives are then removed based on the 2D cross-sections of the vessel-aligned cubes and the ResNet-18 classifier. Huang \textit{et al.} \cite{huang2020penet} presented an end-to-end scalable deep learning model, PENet, which employs volumetric CT imaging to automatically diagnose PE. The CNN-LSTM architecture proposed by Huhtanen \textit{et al.} \cite{huhtanen2022automated} comprises an InceptionResNet v2 CNN architecture and a long short-term memory network that processes the entire CTPA stack as a slice sequence. This architecture demonstrates excellent performance in detecting PE from computed tomography pulmonary angiograms using weakly labeled training data, indicating the potential of deep learning for this task. Shi \textit{et al.} \cite{shi2020automatic} proposed an attention-guided model for the automatic detection of PE, which includes two stages. In the first stage, a 2D convolutional network is trained on a limited set of pixel-level annotated image slices. Patient-level PE predictions are obtained in the second stage. Suman \textit{et al.} \cite{suman2021attention} proposed a CNN-LSTM network based on an attention mechanism, consisting of two parts: a CNN classifier for capturing image attributes and labels, and a sequence model for learning inter-slice dependencies. The CNN extracts features for each slice, and the sequence model with an attention mechanism combines these spatial features and captures long-range dependencies to provide global variation information for each CT slice.

For segmentation tasks, several deep learning-based computer-aided diagnosis research methods have been proposed for segmenting PE in recent years. Yuan \textit{et al.} \cite{yuan2021resd} proposed a ResD-Unet architecture for pulmonary artery segmentation that achieves high accuracy and efficiency by combining the UNet network with novel Residual-Dense blocks and a hybrid loss function, addressing challenges in estimating the severity of PE. Guo \textit{et al.} \cite{guo2022aanet} proposed an artery-aware 3D fully convolutional network, AANet, to detect PE, which addresses the challenge of detecting PEs that appear as dark spots among bright regions of blood arteries in CTPA images. However, due to the complex structure of pulmonary artery layers, the presence of numerous branches, and the large variability in individual sizes and growth directions of each layer, segmenting pulmonary artery embolism detection is more challenging than other lesion detections such as pulmonary nodules. Consequently, selecting a suitable network structure based on the characteristics of the PE CT is a valuable topic that warrants further investigation.

The research methods mentioned above offer valuable references and insights for this paper, providing avenues for further exploration. However, current domestic and international research on the diagnosis of PE predominantly focuses on binary detection of the disease's presence or absence, with limited research on automatic segmentation of PE. Deep neural networks based on U-shaped architecture and skip connections are commonly employed in medical image tasks.  Despite their utility, convolutional networks encounter inherent limitations that hinder their ability to effectively learn global information, particularly when dealing with diseases characterized by numerous small lesions, such as PE. This problem remains unaddressed in both UNet \cite{ronneberger2015u} and UNet++ \cite{zhou2018unet++} network models. Swin-UNet \cite{cao2023swin}, on the other hand, offers a contrast by enabling global semantic feature learning; it achieves this by reintroducing tokenized image blocks into the Transformer-based U-shaped En-Decoder architecture through skip connections. However, Swin-UNet confronts its own set of challenges while trying to achieve detailed segmentation results during the upsampling process. This is due to its exclusive reliance on a pure Transformer structure, with no incorporation of convolutional operations. As a result, the model is prone to interference from other lesions (e.g., nodules), other components of lung tissue, and noise during the process of PE CT image segmentation.

\begin{figure}[h]
    \centering
   \includegraphics[width=0.90\linewidth]{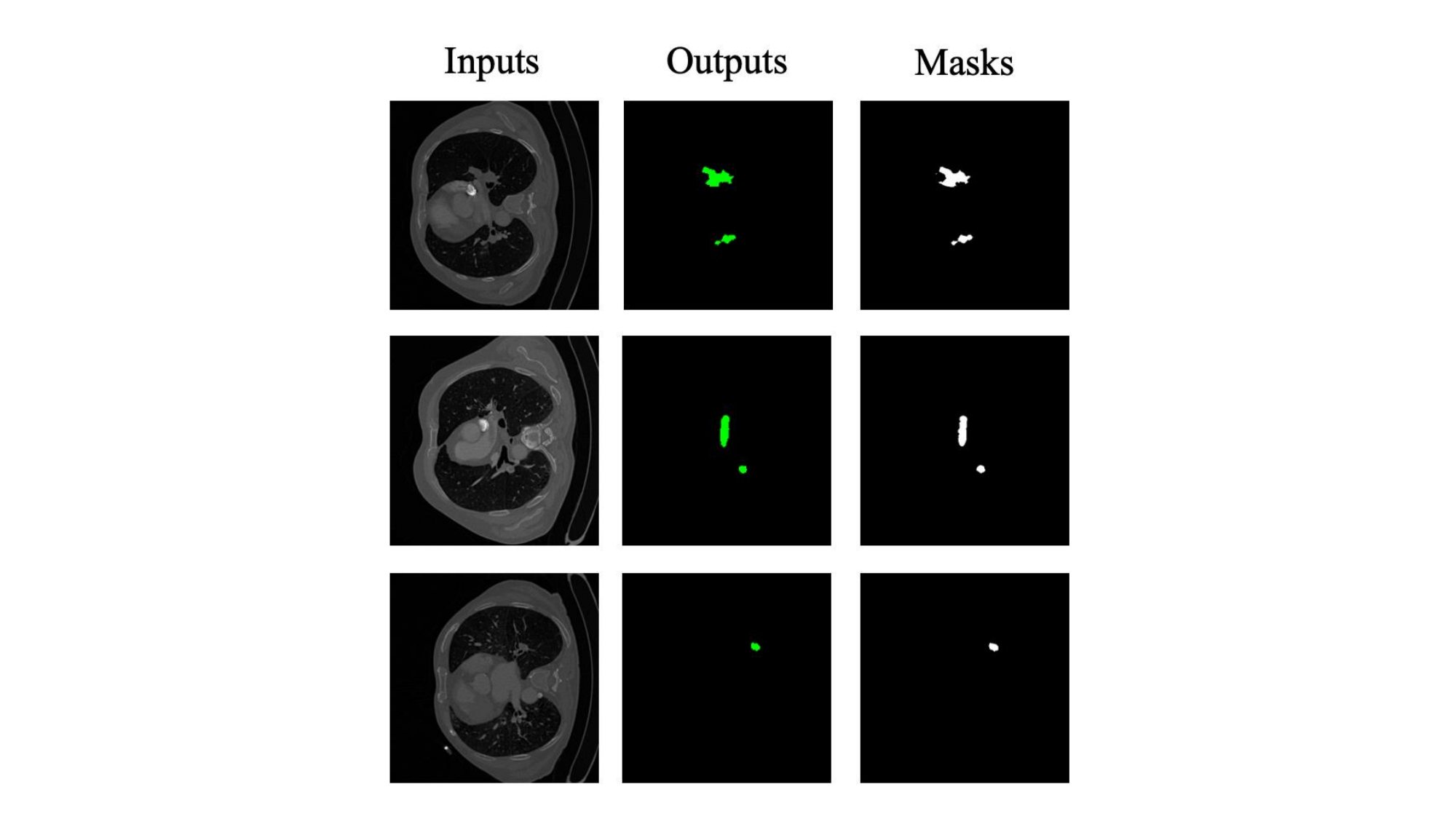}
   \caption{The test results are presented in the following order: from left to right, the input images, the output segmentations, and the ground truth.}
\label{fig:results}
\end{figure}

Based on the findings discussed above, this paper proposes the introduction of a multi-fusion dense skip connection, analogous to UNet++, into the Swin-UNet model structure. This integration enables the effective incorporation of downsampled feature maps of various depths obtained from the Swin-UNet model structure. Consequently, the model can concentrate on both contextual information and spatial visual information. Moreover, feature maps of different depths can partially share the same encoder and perform joint feedback learning through depth supervision. To compensate for the lack of local spatial feature extraction in the Swin Transformer module at the bottleneck layer, this paper also proposes replacing the original bottleneck layer of the Swin-UNet with a convolutional module. The hybrid architecture model is capable of effectively extracting local spatial features of the image alongside overall global features and combining deep and shallow semantic information to accurately identify the segmentation of PE, as illustrated in Fig. \ref{fig:results}.

\section{Method}

\subsection{Overview of the Architecture}
\begin{figure*}[t]
    \centering
   \includegraphics[width=0.80\linewidth]{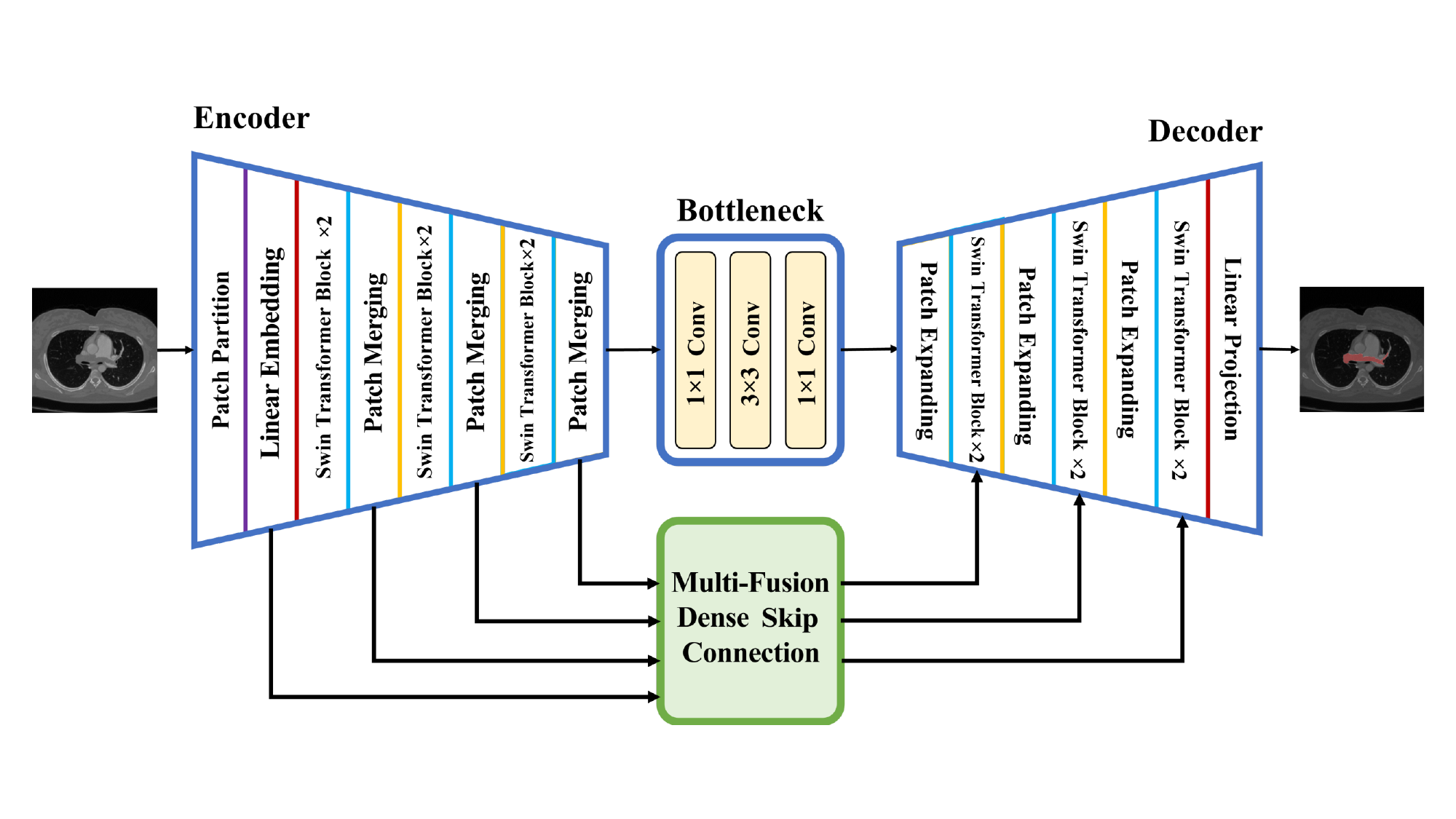}   \caption{Overall structure of the network model. This method incorporates multiple fusion dense skip connections between the encoder and decoder, utilizing the Swin Transformer as the encoder. Additionally, we use CNN in bottleneck and Multi-Fusion Dense Skip Connections to make up for the Transformer's shortcomings in local spatial feature extraction.}
\label{fig:Overall}
\label{fig:two}
\end{figure*}

The overall architecture of the SCUNet++ model proposed in this paper is illustrated in Fig. \ref{fig:Overall}. The model comprises four main modules: an encoder module, a CNN Bottleneck module, a decoder module, and a multi-fusion dense skip connection module. For the encoder module, the image is transformed into a sequence embedding, and the CT slice with a size of W$\times$H is divided into non-overlapping blocks of size $4\times4$. Then, the feature dimension is projected to the C dimension by a linear embedding layer. The transformed patches are then passed through a series of Swin Transformer blocks and a patch merging layer to generate hierarchical feature representations. The patch merging layer is responsible for downsampling and dimension increase, while Swin Transformer blocks are responsible for feature representation learning. After the encoder, the feature map is input to the CNN bottleneck module, which employs CNN blocks for alternative construction to compensate for the Swin Transformer blocks' deficiency in local spatial feature extraction in the image. The multi-fusion dense skip connection module, inspired by the UNet++ skip connection method, is added between the decoder and the encoder, based on Swin Transformer blocks. It fuses the extracted contextual features with multi-scale features from the encoder through multi-fusion dense skip connections to compensate for the loss of spatial information caused by downsampling. The decoder module consists of a Swin Transformer block and a patch-expanding layer. The patch expanding layer reshapes feature maps of adjacent dimensions into larger feature maps with a resolution of $2\times$ upsampling, while the Swin Transformer block is responsible for feature representation learning. Finally, the feature maps' resolution is restored to the input resolution (W$\times$H) by 4-fold upsampling using the last patch expanding layer. A linear projection layer is then applied to this feature map to perform pixel-level segmentation prediction.

\subsection{Double Swin-Transformer Block}
\begin{figure}[H]
    \centering
   \includegraphics[width=0.90\linewidth]{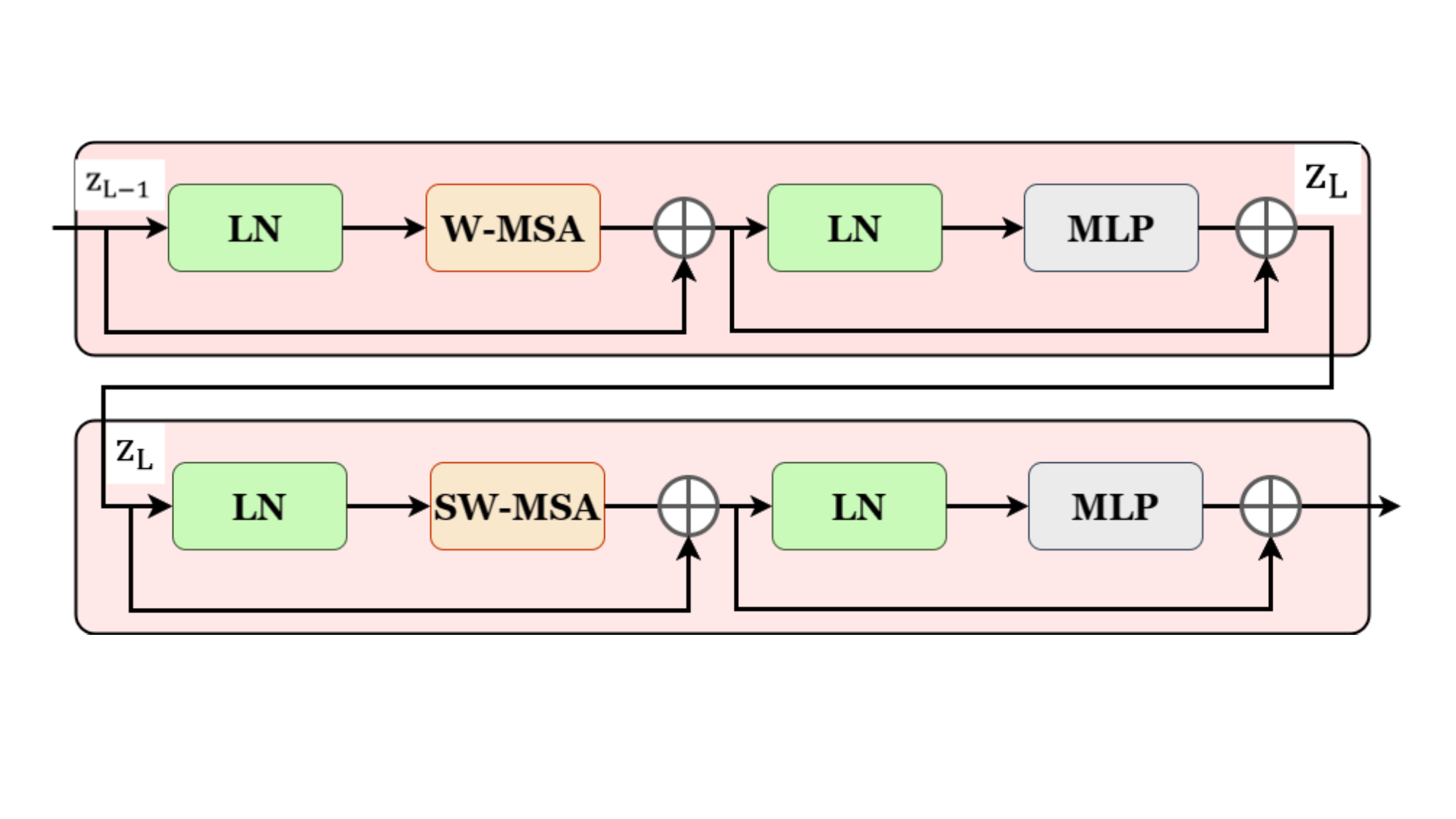}
   \caption{Swin-Transformer module. MSA denotes the multiheaded attention module and MLP represents the multilayer perceptron module.}
\label{fig:Transformer}
\end{figure}

As illustrated in Fig. \ref{fig:Transformer}, the Swin-Transformer block is constructed based on a shifted window, which sets it apart from the conventional Multihead Self-Attention (MSA) module. The Double Swin-Transformer block proposed in this paper comprises two consecutive Swin-Transformer blocks. Each Swin-Transformer block includes an LN layer, an MSA module, a residual connection, and a two-layer MLP with a GELU nonlinear activation function. Moreover, we apply the window-based multi-headed self-attention (W-MSA) module and the shifted-window-based multi-headed self-attention (SW-MSA) module to these two consecutive Swin-Transformer blocks, respectively. This approach enhances the model's ability to capture long-range dependencies and learn global and remote semantic information interactions.

\subsection{Encoder}
The encoder is built based on Double Swin-Transformer blocks. Firstly, the image is transformed into a sequence embedding, and the patch partition divides the W$\times$H$\times$3 slice into non-overlapping blocks of size 4$\times$4, so the feature dimension of each block becomes $4\times4\times3=48$. The linear embedding layer projects the feature dimension onto C. Then, the feature of size H/4$\times$W/4$\times$C is used as the input for two consecutive Swin-Transform blocks to learn feature representation. At the last, downsampling by a factor of 2 is performed via the patch merging layer to reduce the number of representations and double the feature dimension. So, the changes in dimension and resolution are completed at the patch merging layer. This process is repeated three times in the encoder, each of them comprising two Swin-Transformer blocks and one patch merging layer.

\subsection{Patch Merging Layer}
When combining pixel points based on the parity of rows and columns, there are four possible combinations. So, the input patch can be divided into four parts, and use the patch merging layer to merge these four parts. This reduces the feature resolution to half of the original resolution. Since the merging operation increases the feature dimension to four times the original dimension, a linear layer is applied afterward to reduce the feature dimension to double the original dimension.

\subsection{Bottleneck (CNN Block)}
Using only Swin-Transformer blocks to construct the bottleneck for extracting local spatial features in images may be insufficient. Therefore, we attempt to use CNN blocks to replace the original bottleneck to fully learn deep feature representations, and make up for the deficiency of the network in extracting local spatial features in images. The CNN block comprises a network structure that utilizes a $1\times1Conv-3\times3Conv-1\times1Conv$ sequence, with a BN-ReLU operation applied before each convolution. In the bottleneck, the feature size and resolution remain unchanged to ensure consistency in feature representation.

\subsection{Decoder}
The structure of the decoder corresponds to the encoder. In the decoder, the patch expanding layer performs 2x upsampling on the extracted deep features and reshapes the adjacent dimension feature maps into higher-resolution feature maps. Accordingly, the feature dimension is reduced to half of the original dimension.

\subsection{Patch Expanding Layer}
Taking the first patch expansion layer as an exemplar, a linear layer is employed on the input features (W/32$\times$H/32$\times$8C) preceding the upsampling process. This action aims to amplify the feature dimension, doubling its original value (W/32$\times$H/32$\times$16C). Subsequent to this, the resolution of the input features is enhanced to twice the initial input resolution via the rearrange operation. Conversely, the feature dimension decreases to a quarter of the initial input feature dimension.

\subsection{Multi-Fusion Dense Skip Connection}
\begin{figure}[h]
\centering
   \includegraphics[width=0.90\linewidth]{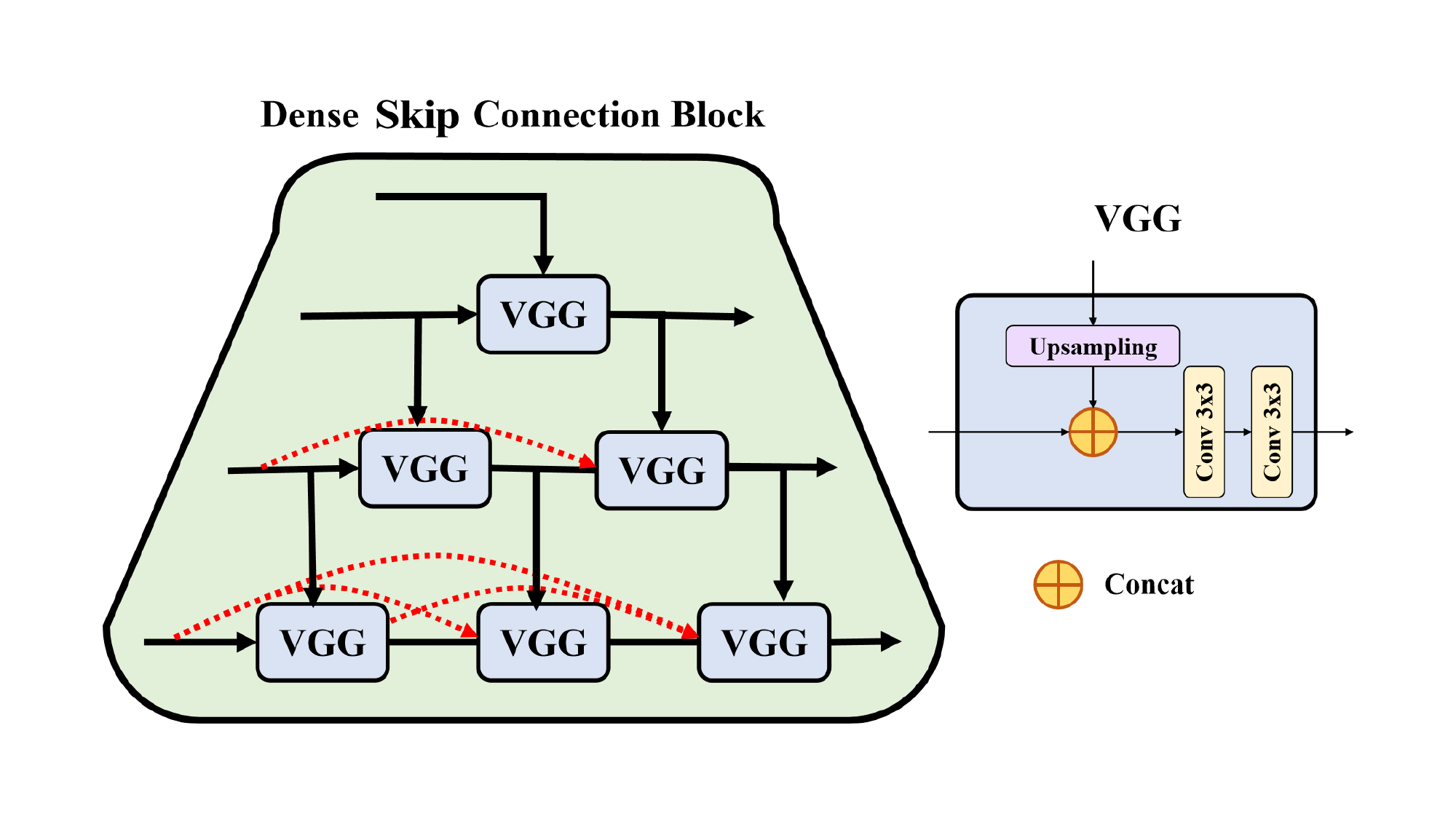}
   \caption{Multi-Fusion Dense Skip Connection module.}
\label{fig:Multi-fusion}
\end{figure}

Inspired by the skip connection concept of UNet++, the SCUNet++ uses a multi-fusion dense skip connection approach to attain aggregated semantically scaled and highly flexible encoder features within the decoder sub-network. As illustrated in Fig. \ref{fig:Multi-fusion}, the proposed model is designed with dense connections, incorporating both long and short connections by introducing nested and dense multi-fusion dense skip connections. Moreover, a convolutional structure is added to fill the hollow spaces in the center of the model to compensate for the lack of image feature extraction detail resulting from the pure Transformer structure of the Swin-UNet. This strategy effectively reduces the semantic gap between the encoder and decoder while capturing various levels of features and different sizes of perceptual fields to obtain richer multi-scale information. Consequently, our model can efficiently extract local spatial features and overall global features of CTPA images, integrating deep and shallow semantic information and performing stitching processing to minimize the loss of spatial information caused by the downsampling process.

\section{Experiment}

\subsection{Dataset}
\begin{figure}[h]
\centering
   \includegraphics[width=0.90\linewidth]{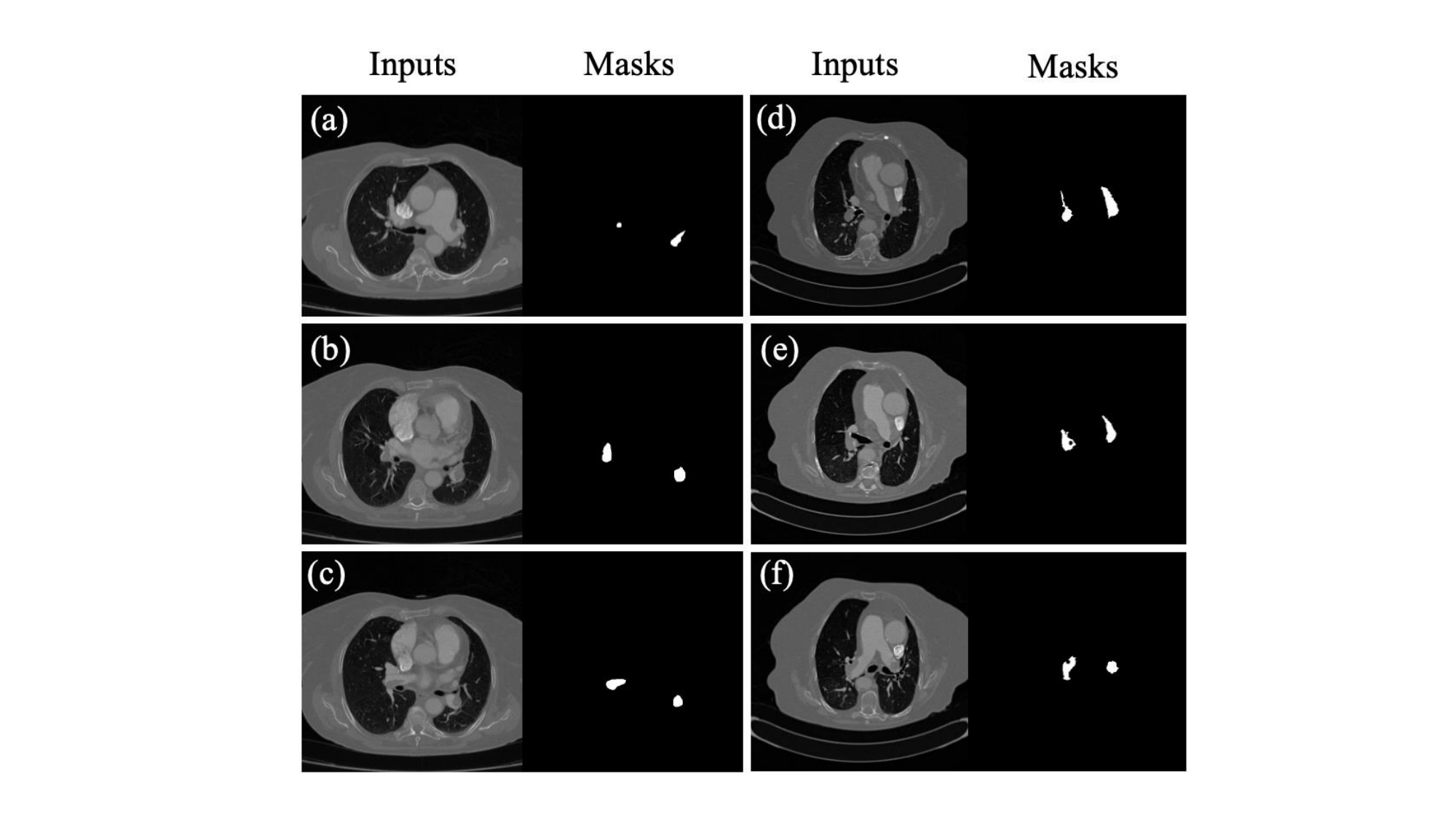}
   \caption{Original PE dataset. Figures (a-c) present examples from the CAD-PE dataset, while figures (d-f) showcase examples from the FUMPE dataset.}
\label{fig:dataset}
\end{figure}

As depicted in Fig. \ref{fig:dataset}, we conducted experiments using publicly available datasets FUMPE \cite{masoudi2018new} and CAD-PE \cite{gonzalez2020computer}. We only selected fixed-angle PE images and corresponding labels for each case. Owing to inaccuracies in the initial annotation results of the FUMPE dataset, we collaborated with medical professionals to scrutinize and re-annotate the dataset. Following a thorough review process, we ultimately secured 8,792 CTPA images obtained from 35 patients, complete with revised annotations provided by the physicians. The CAD-PE dataset contains 91 computed tomography pulmonary angiography images with positive PE, with all thrombi segmented by experienced radiologists in each scan. After the screening, the CAD-PE dataset consists of 8,900 CTPA images of PE. Simultaneously, we standardized the image size of both datasets to 512 $\times$ 512 pixels and divided each PE case into a 9:1 ratio for training and testing.

\subsection{Implementation Details}
In this study, we utilized the PyTorch framework to construct the SCUNet++ network model and trained it using an NVIDIA GeForce RTX 3090 GPU. Specifically, we set the batch size to 12 and employed the Adam optimization algorithm to update the model's gradient. The learning rate was set at 0.0001, the number of epochs at 150, and the training time at 23 hours. To evaluate the model's performance in image segmentation, we assessed the segmentation accuracy using the Dice similarity coefficient (DSC) \cite{milletari2016v} and 95\% Hausdorff distance (HD95) \cite{kim2015quantitative}, both of which are widely applied in image segmentation. Given the predicted segmentation mask as $X$ and the ground-truth label as $Y$, the DSC is calculated as:
\begin{equation}
DSC=\frac{2|X\cap Y|}{|X|+|Y|},
\end{equation}
where $|X|$ and $|Y|$ represent the area of the segmented result and the label while $|X\cap Y|$ refers to the area of the overlapping part of the segmented result and the label.

Analogously, the formula for HD95 can be derived as follows:
\begin{equation}
\begin{split}
&d_H(X,Y) =\max\left\{d_{XY},d_{YX}\right\} \\
&=\max\left\{\max\limits_{x\in X}{\min\limits_{y\in Y}{d(x,y)}},\max\limits_{y\in Y}{\min\limits_{x\in X}{d(x,y)}}\right\},
\end{split}
\end{equation}
where $X$ and $Y$ denote the segmented pixel values and labeled pixel values, respectively, and $d(x,y)$ represents the Euclidean distance between pixel values $X$, $Y$.

The HD95 value is obtained by multiplying the final value by 0.95, with the intention of excluding some unreasonable distances from the cluster points and maintaining overall distance stability. The HD95 value represents the relative distance size between two sample boundary values; a smaller distance value indicates a closer prediction map and segmentation map boundary, leading to more accurate segmentation. DSC measures the overlap between the segmentation result and the label, while HD95 quantifies the maximum distance between the network prediction region and the label. Consequently, a higher DSC and lower HD95 indicate better performance of the semantic segmentation model. DSC is more sensitive to the internal and true regions of the segmentation area, whereas HD95 primarily focuses on the boundary of the segmentation result. By combining these two metrics, we can objectively and quantitatively evaluate the segmentation performance of the model.

\subsection{Comparison with Typical Segmentation Models}

\begin{table}[h]
\caption{Comparison of DSC and HD95 for various network models utilizing the FUMPE dataset.}
\label{FUMPE}
\centering
    {\small{
\begin{tabular}{llr}
\toprule
Methods & DSC(\%) & HD95 \\
\midrule
UNet\cite{ronneberger2015u} & $78.13\pm 13.87$ & $6.86\pm 15.81$  \\
UNet++\cite{zhou2018unet++} & $77.16\pm 16.25$ & $5.80\pm 4.57$ \\
Swin-UNet\cite{cao2023swin} & $60.80\pm 28.49$ & $20.20\pm 28.93$ \\
ResD-UNet\cite{yuan2021resd} & $76.48\pm 20.12$ & $22.25\pm 35.42$ \\
\textbf{SCUNet++(Ours)} & \textbf{83.47$\pm$ 5.57} & \textbf{3.83$\pm$ 1.02} \\
\bottomrule
\end{tabular}
}}
\end{table}

\begin{table}[h]
\caption{Comparison of DSC and HD95 for various network models utilizing the CAD-PE dataset.}
\label{CAD}
\centering
    {\small{
\begin{tabular}{llr}
\toprule
Methods & DSC(\%) & HD95 \\
\midrule
UNet & $73.79\pm 20.60$ & $6.86\pm 24.38$  \\
UNet++ & $77.48\pm 19.16$ & $13.80\pm 33.12$ \\
Swin-UNet & $67.49\pm 24.13$ & $16.56\pm 36.13$ \\
ResD-UNet & $73.58\pm 22.21$ & $27.25\pm 47.24$ \\
\textbf{SCUNet++(Ours)} & \textbf{83.42$\pm$ 6.12} & \textbf{5.10$\pm$ 9.14} \\
\bottomrule
\end{tabular}
}}
\end{table}

\begin{figure*}[t]
\centering
   \includegraphics[width=0.80\linewidth]{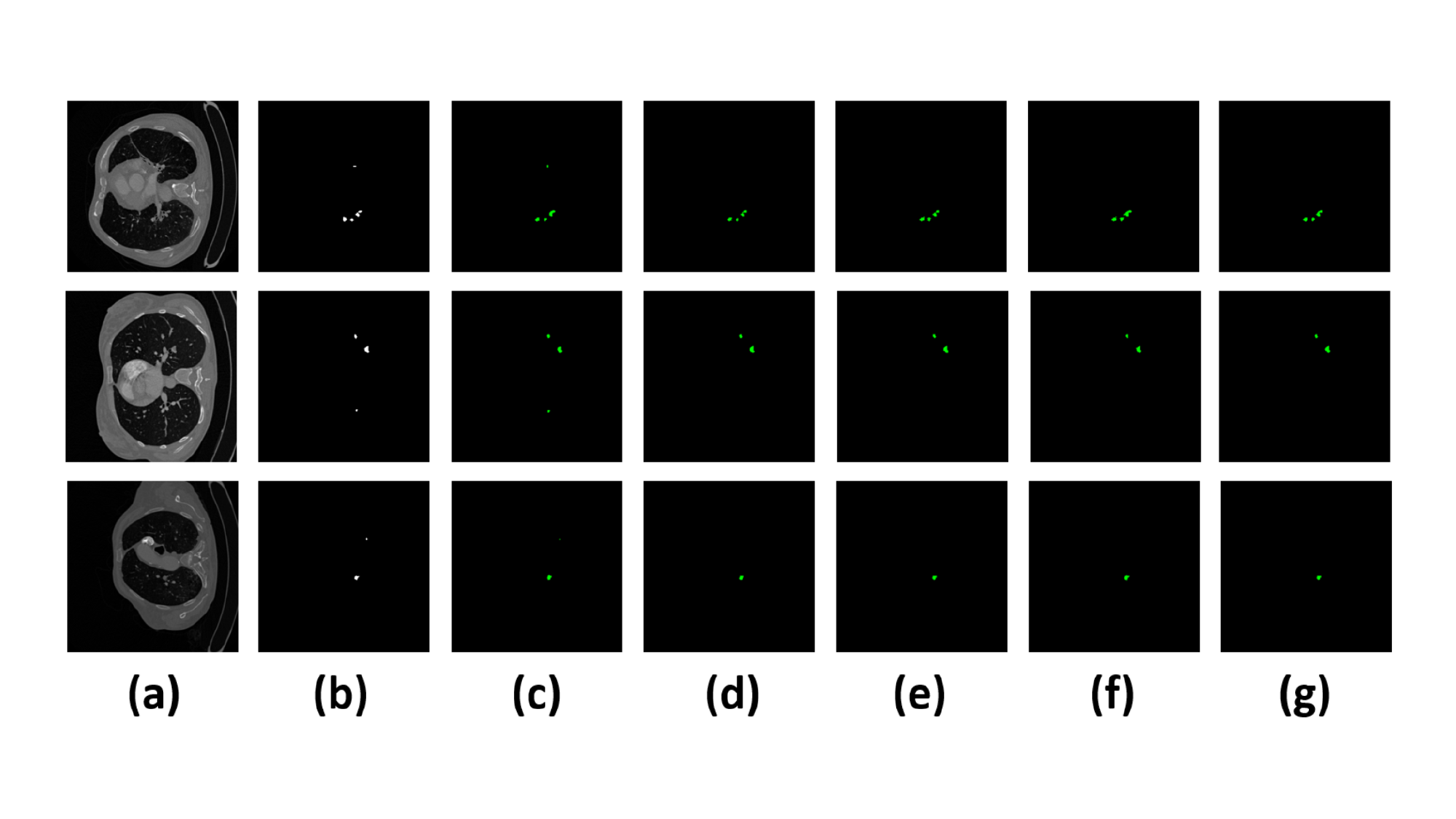}
   \caption{Comparison of segmentation performance of different network models on the CAD-PE dataset: (a) input image; (b) ground truth mask; (c) the proposed SCUNet++ model; (d) UNet++ model; (e) UNet model; (f) Swin-UNet model; and (g) ResD-UNet model.}

\label{fig:Comparison}
\label{fig:Comparison2}
\end{figure*}

In the experiment, we compared the segmentation performance of SCUNet++ with some traditional or state-of-the-art segmentation models. Fig. ~\ref{fig:Comparison} shows the segmentation results of our proposed SCUNet++ and classical segmentation models. To ensure the reliability of the results, we have done experiments on both FUNMPE and CAD-PE datasets and obtained similar results. The results presented in Table \ref{FUMPE} and Table \ref{CAD} indicate that SCUNet++ achieves the best performance in segmenting CT images of PE. 

In the CAD-PE dataset, when compared to the UNet and UNet++, the DSC for SCUNet++ improves from 73.79\% and 77.48\% to 83.42\%, while the HD95 decreases from 6.86 and 13.80 to 5.10. This improvement can be attributed to the Swin Transformer module within the SCUNet++ network model, which is based on the attention mechanism. This module can capture global connections in one step while concentrating on local connections between elements, thereby enhancing the segmentation performance.

Furthermore, we compared the SCUNet++ with the Swin-UNet. The results showed that the DSC of SCUNet++ improved from 67.49\% to 83.42\%, and HD95 decreased from 16.56 to 5.10 when applied to the CT dataset of PE. This improvement is due to the SCUNet++ addressing the limitations of a pure Transformer in extracting detailed features from medical images. By deepening the network structure and adding a multi-scale feature fusion module to each Swin Transformer block, the SCUNet++ model effectively fuses features from multiple scales, thus improving segmentation performance. 

And, we also compared the SCUNet++ with the state-of-the-art segmentation model ResD-UNet. The results showed that the DSC of SCUNet++ improved from 73.58\% to 83.42\%, and HD95 decreased from 27.25 to 5.10 when applied to the CT dataset of PE. 

At the same time, we repeated comparative experiments on the FUMPE dataset, and the experimental results were similar to those on the CAD-PE dataset, further proving the good generalization. When compared to the UNet and UNet++, the DSC for SCUNet++ improves from 78.13\% and 77.16\% to 83.47\%, while the HD95 decreases from 6.86 and 5.80 to 3.83. When compared to the Swin-UNet and ResD-UNet, the DSC for SCUNet++ improves from 60.80\% and 76.48\% to 83.47\%, while the HD95 decreases from 20.20 and 22.25 to 3.83.

In conclusion, we compared the model parameters, training time, and inference time of SCUNet++ with other advanced methods. As demonstrated in Table \ref{tabExtra}, our proposed method exhibits a less significant increase in model parameters when tested on the FUMPE dataset (akin to CAD-PE) while maintaining a notable advantage in terms of training and inference speed. However, within real-world scenarios pertaining to CT image segmentation tasks of pulmonary embolisms, accuracy is generally prioritized over speed and computational efficiency. Thus, the superiority of our proposed model is underscored by its substantial improvement in the accuracy of pulmonary embolism CT image segmentation, compared to other contemporary state-of-the-art models.

\begin{table}[h]
\caption{Comparison of model parameters, training, and inference time for various network models.}
\label{tabExtra}
\centering
    {\small{
\begin{tabular}{p{0.3\linewidth}p{0.09\linewidth}p{0.21\linewidth}p{0.2\linewidth}}
\toprule
Methods & Params (M) & Training time (min/epoch) & Inference time (s/it) \\
\midrule
UNet\cite{ronneberger2015u} & $32.93$ & $13.4046$ & $0.2045$ \\
UNet++\cite{zhou2018unet++} & $34.96$ & $13.5667$ & $0.2058$ \\
Swin-UNet\cite{cao2023swin} & $25.91$ & $11.5383$ & $0.2016$ \\
ResD-UNet\cite{yuan2021resd} & $50.83$ & $13.3416$ & $0.2035$ \\
\textbf{SCUNet++ (Ours)} & \textbf{60.11} & \textbf{12.2805} & \textbf{0.2020} \\
\bottomrule
\end{tabular}
}}
\end{table}

\subsection{Ablation Study}
\begin{figure*}[h]
\centering
   \includegraphics[width=0.80\linewidth]{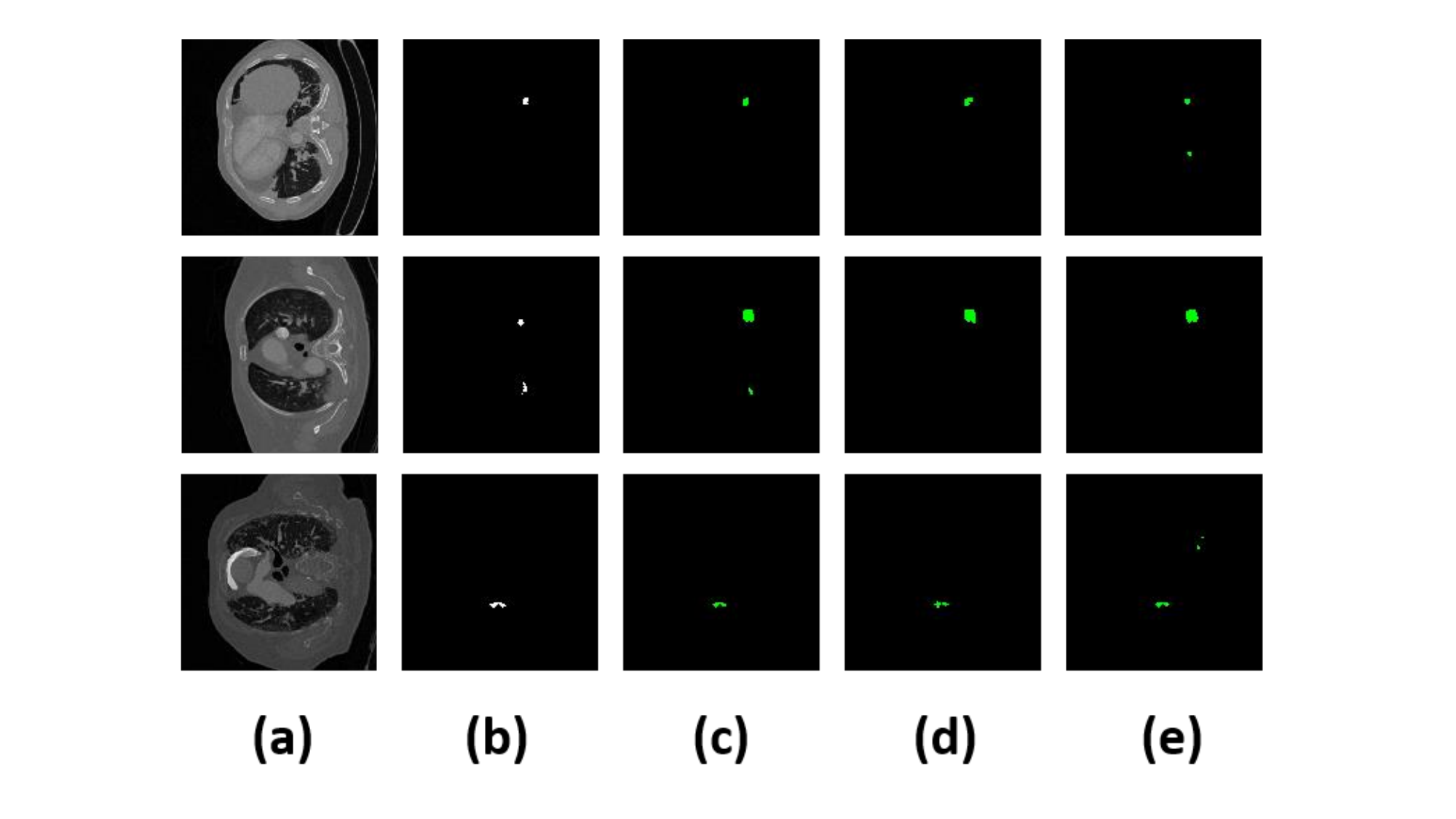}
   \caption{Comparison of segmentation performance of different network models on the CAD-PE dataset: (a) input image; (b) ground truth mask; (c) the proposed SCUNet++ model; (d) SCUNet(w/o Dense skip connection); (e) SUNet++(w/o CNN bottleneck block).}
\label{fig:Ablation}
\end{figure*}

In this section, as shown in Fig.~\ref{fig:Ablation}, we assess the impact of skip connections and CNN blocks on the model's segmentation performance through ablation experiments. As illustrated in Table \ref{xr_FUMPE} and Table \ref{xr_CAD}, we evaluate each module's influence on the overall model by comparing the Dice values on the test set for the following configurations: without(w/o) Dense skip connection, w/o CNN block, and the complete SCUNet++ model.

In the CAD-PE dataset, when we added a dense skip connection module on the basis of SCUNet, the DSC increased from 67.15\% to 83.41\%, while the HD95 decreased from 22.83 to 5.10. This indicates that by removing the dense skip connection module, the segmentation effect is unsatisfactory due to the semantic gap between the encoder and decoder. When we added a CNN bottleneck block module on the basis of SUNet++, the DSC increased from 75.27\% to 83.41\%, while the HD95 decreased from 18.01 to 5.10. This indicates that by removing the CNN bottleneck block, the extraction of features in the image's local space is insufficient because only the Swin-Transformer block is used to construct the bottleneck, which will lead to the lack of feature extraction in the local space of the image.

At the same time, we also conducted the same ablation experiment on the FUMPE dataset and obtained similar results. SCUNet++ achieved the best results in both DSC and HD95, 83.47 and 3.83. Based on the experimental results on the two datasets, we can conclude that both individual modules have a significant impact on the model's segmentation results. 

\begin{table}[h]
\caption{Comparison of ablation experiments utilizing FUMPE dataset.}
\centering
  \label{xr_FUMPE}

    {\small{
\begin{tabular}{llr}
\toprule
Methods & DSC(\%) & HD95 \\
\midrule
SCUNet(w/o Dense skip connection) & 80.52 & 4.25  \\
SUNet++(w/o CNN bottleneck block) & 80.87 & 3.98 \\
\textbf{SCUNet++(Ours)} & \textbf{83.47} & \textbf{3.83} \\
\bottomrule
\end{tabular}
}}
\end{table}

\begin{table}[h]
\caption{Comparison of ablation experiments utilizing CAD-PE dataset.}
  \label{xr_CAD}

\centering
    {\small{
\begin{tabular}{llr}
\toprule
Methods & DSC(\%) & HD95 \\
\midrule
SCUNet(w/o Dense skip connection) & 67.15 & 22.83  \\
SUNet++(w/o CNN bottleneck block) & 75.27 & 18.01 \\
\textbf{SCUNet++(Ours)} & \textbf{83.41} & \textbf{5.10} \\
\bottomrule
\end{tabular}
}}
\end{table}

The SCUNet++ network model addresses these issues by strengthening connections through the introduction of nested multi-fusion dense skip connections, ultimately designing dense connections with both long and short connections. Additionally, it incorporates a convolutional structure to fill the hollow position in the middle of the model. This structure enables the capture of features at different levels and perceptual fields of varying sizes, obtaining richer multi-scale information. Furthermore, SCUNet++ employs CNN blocks as a bottleneck to fully learn deep feature representation, compensating for the network's deficiency in extracting local spatial features from images.

Therefore, the ablation experiments confirm our hypothesis that the inclusion of skip connections and CNN blocks improves the network model's segmentation Dice values. The fusion of these modules allows the proposed network model to effectively utilize local spatial features, overall global features, and deep and shallow semantic information, ultimately greatly enhancing the accuracy and stability of PE segmentation.

\section{Conclusion}
This paper proposes a novel automatic segmentation network, SCUNet++, for CTPA images. We adjust the network structure based on an analysis of the disease characteristics and the actual segmentation processes. This network introduces multiple fused dense skip connections between the encoder and decoder, allowing the decoder sub-network to fuse features of different scales and compensate for the spatial information loss caused by downsampling. Our experimental results on the published FUMPE and CAD-PE datasets demonstrate that this method outperforms other state-of-the-art methods in segmentation accuracy, positively impacting the diagnosis of PE.

In the future, we plan to use more reliable annotated PE data to further improve the network's accuracy and robustness, ensuring the method's stability and reliability. Additionally, we will continue to collaborate with PE specialists and include additional patient information, such as age, gender, surgical history, and other relevant factors, to develop quantifiable and reliable evaluation guidelines. These guidelines will assist doctors in better assessing patients' conditions and risk levels, ultimately promoting the provision of more accurate treatment plans for patients.


\end{document}